\documentstyle[preprint,aps,epsf]{revtex}
\begin{document}

\draft
\title{Cosmic Microwave Background non--Gaussian signatures\\
from analytical texture models}

\author{Alejandro Gangui$^{1,2}$ and Silvia Mollerach$^{2}$}
\address{$^{1}$ICTP -- International Center for Theoretical Physics,\\
P.~O.~Box 586, 34100 Trieste, Italy.}
\address{$^{2}$SISSA -- International School for Advanced Studies,\\
Via Beirut 2--4, I--34013 Trieste, Italy.}

\maketitle
\tighten
\begin{abstract}
Using an analytical model for the Cosmic Microwave Background  
anisotropies produced by textures, we compute the resulting 
collapsed three--point correlation function and the {\it rms} expected 
value due to the cosmic variance. 
We apply our calculations to the 
{\sl COBE}--DMR experiment and test the consistency of the model 
with the observational results.
We also show that an experiment with smaller 
angular resolution can put bounds to the model parameters.
\end{abstract}
\narrowtext

\section{Introduction}

An important problem in cosmology is that of finding out the
source of the density perturbations leading to the large scale structure
formation. Competing models are  inflationary scenarios and
topological defects, these latter formed as the consequence
of a symmetry breaking phase transition in the early universe.

A possible way of trying to discriminate between models is
given by the anisotropies in the Cosmic Microwave Background 
(CMB) radiation predicted by them.
In inflationary models, large scale anisotropies arise from the 
Sachs--Wolfe effect \cite{sa67} and are proportional to the
fluctuations of the 
gravitational potential on the last--scattering surface. These are 
generated as a result of quantum fluctuations of scalar fields 
during inflation. The resulting anisotropies are consistent with the 
analysis of the two--point correlation function of the two--year 
{\sl COBE}--DMR data \cite{be94}, which fixes the amplitude of the
power spectrum and constrains the spectral index, provided that the
couplings of the inflaton field are very weak. As a consequence, the
CMB anisotropies produced in inflationary models are nearly Gaussian 
distributed. 

Defect--induced large scale structure formation has been well studied
recently \cite{vi94,pe94,ru95,be93} 
and its predictions confront successfully 
against a large bulk of observational data. 
The amplitude of the anisotropies, which can be determined
using the {\sl COBE}--DMR data, is directly related to
the symmetry breaking energy scale and corresponds to 
a Grand Unification Theory (GUT) phase transition.
However, the production of anisotropy maps from numerical
simulations of the field equations
turns out to be a fabulous task, and therefore some analytical
insight is desirable.
This is the reason why people turned to consider
analytical (although simplified) models for CMB anisotropies generated
by defects, wherein the physics is more transparent and computations
can be pursued straightforwardly \cite{pe93,ma95,ga95}.
Although it is clear that full range numerical simulations will have
the last word regarding these and other observable signatures
of defects, in the meantime progress with these simplified models
can hint some of their main features.

CMB maps contain more  information than the one that can be extracted
by studying just the two--point correlation function. Recently,
Hinshaw et al. \cite{hi95} analysed the three--point function of the
two--year {\sl COBE}--DMR data and found a non--vanishing signal, 
although consistent with the level of cosmic
variance associated to a Gaussian process. The predictions for the
three--point CMB function in inflationary models have recently been
worked out in detail. 
The contribution coming from the non--linearities in the inflaton 
evolution and in the relation between the inflaton fluctuations and 
the CMB resulting anisotropies have been computed in Refs. 
\cite{fa93,ga94,ga94b}. Also the post recombination mildly 
non--linear evolution of the perturbations gives rise to a 
non--vanishing contribution through the Rees--Sciama effect
\cite{lu93,mo95,mu95}. Both contributions result to be more than 
three orders of magnitude smaller than the associated cosmic variance, 
and are thus consistent with the {\sl COBE}--DMR data.
It is interesting to know whether the predictions of topological defect
models are also consistent with the {\sl COBE}--DMR three--point
function data or not.
As defects are the typical example of non--Gaussian perturbation
sources, the resulting three--point function and its cosmic variance
can easily differ from the Gaussian ones. In this paper we apply 
a recently proposed analytical model of anisotropies produced by
textures \cite{ma95} to study the predictions for the three--point 
correlation function and cosmic variance of the CMB 
and we compare this theoretical band with the {\sl COBE}--DMR data.

As an example that this statistics 
can be a useful test for models of structure formation, let us note
that, based on the analysis of the three--point function of the 
{\sl COBE}--DMR data, Hinshaw et al. \cite{hi95} were able to rule 
out an inflationary model for primordial isocurvature baryon (PIB)
fluctuations proposed by Yamamoto and Sasaki \cite{ya94}. In this
model the predicted mean skewness vanishes and the {\it rms} skewness
(cosmic variance) results several orders of magnitude below the
standard adiabatic models one, and the authors suggest that the 
{\it rms} value of the full three--point function would be tiny as well.
This would make the theoretical error bars for the three--point
function predicted by this PIB model smaller that the actual data from
the maps, hence running into conflict with the result of the 
{\sl COBE}--DMR analysis. 

Likewise, it is interesting to check whether the non--Gaussian
signatures predicted by the analytical texture model are in agreement with
the analyses of the maps, and if not what the constraints are. This is
the main aim of the present paper. 

\section{The collapsed three--point correlation function}

The three--point correlation function for points at three arbitrary
angular separations $\alpha$, $\beta$ and $\gamma$ is given by the
average product of temperature fluctuations in all possible three
directions with those angular separations among them \cite{ga94}. 
In this paper, for
simplicity, we will restrict ourselves to the collapsed case,
corresponding to the choice $\alpha=\beta$ and $\gamma=0$, that is one
of the cases analysed for the {\sl COBE}--DMR data \cite{hi94,hi95}
(the other is the equilateral one,
$\alpha=\beta=\gamma$). The collapsed three--point correlation
function of the CMB is given by
\begin{equation}
C_3(\alpha) \equiv \int \frac{d\Omega_{\hat \gamma_1}}{4\pi}
\int \frac{d\Omega_{\hat \gamma_2}}{2\pi}
\Delta T  (\hat\gamma_1) \Delta T^2 (\hat\gamma_2)
\delta(\hat\gamma_1\cdot\hat\gamma_2 -\cos \alpha).
\label{skew1}
\end{equation}
For $\alpha=0$, we recover the well--known expression for the
skewness, $C_3(0)$. By expanding the temperature fluctuations in
spherical harmonics
${\Delta T \over T} (\vec x , \hat\gamma) = \sum_{\ell,m} a_{\ell}^{m}(\vec x)
Y_{\ell}^{ m} (\hat\gamma)$, we can write the collapsed
three--point function as
\begin{equation}
\label{coll}
C_3(\alpha) =\frac{T_0^3}{4\pi} \sum_{\ell_1,\ell_2,\ell_3,m_1,m_2,m_3}
P_{\ell_1}(\cos\alpha)
a_{\ell_1 }^{m_1} a_{\ell_2}^{ m_2} a_{\ell_3}^{ m_3}
{\cal W}_{\ell_1} {\cal W}_{\ell_2} {\cal W}_{\ell_3}
\bar {\cal H}_{\ell_1 \ell_2 \ell_3}^{m_1 m_2 m_3} ~,
\end{equation}
where $T_0=2.726 \pm 0.01$K is the mean temperature of the CMB radiation
\cite{ma94},
${\cal W_\ell}$ represents the window function of the particular
experiment and we follow the notation in Ref. \cite{ga94}, defining
\begin{equation}
\label{ana91}
\bar {\cal H}_{\ell_1 \ell_2 \ell_3}^{m_1 m_2 m_3}
\equiv \int
d\Omega_{\hat\gamma} Y_{\ell_1}^{m_1} (\hat\gamma)
Y_{\ell_2}^{m_2}(\hat\gamma) Y_{\ell_3}^{m_3} (\hat\gamma) ~,
\end{equation}
which have a simple expression in terms of Clebsh--Gordan coefficients.

Predictions from different models usually come as
expressions for the ensemble average of the angular bispectrum
$\bigl\langle a_{\ell_1 }^{m_1} a_{\ell_2}^{ m_2} a_{\ell_3}^{ m_3} 
\bigr\rangle$, from which we obtain the predicted $\langle C_3(\alpha)
\rangle$. This corresponds to the mean value expected in an ensemble
of realizations. However, as we can observe just one particular 
realization, we have to take into account the spread of the
distribution of the three--point function values when comparing
a model prediction with the observational results. This is the 
well--known cosmic variance problem \cite{sc91,sr93}. We can estimate
the range of expected values about the mean by the  
{\it rms} dispersion
\begin{equation}
\sigma_{CV}^2(\alpha ) \equiv
\bigl\langle C_3^2 (\alpha ) \bigr\rangle - \bigl\langle C_3 (\alpha )
\bigr\rangle^2.
\label{cv}
\end{equation}

\section{CMB anisotropies from Textures}

CMB anisotropies in defect theories are produced during 
the photon travel 
all the way from the last scattering surface to here. The 
computation is quite involved because it requires simulations 
following the defects, matter and photon evolution over a large number 
of expansion times \cite{pe94,ru95,be93,co94}. 
In the case of textures, the 
non--linear evolution of the scalar field responsible for texture 
formation is the source of the CMB anisotropies. As long as the 
characteristic length scale of these defects is larger
than the Hubble radius $H^{-1}$ the gradients of the field
are frozen in and cannot affect the surrounding matter
components. 
However, when the size of the texture becomes comparable to 
$H^{-1}$
(this latter grows linearly with cosmic time),
microphysical processes can `push' the global fields
so that they acquire trivial field configurations,
reduce their energy (which is radiated away as Goldstone bosons) and,
in those cases where the field was wounded in knots, they unwind.
In this process perturbations in the spacetime metric are 
generated, and hence also the photon geodesics are affected.
These unwinding events are mainly characterised by one length scale,
that of the Hubble radius  
at the moment of collapse, implying that the
resulting perturbations arise at a fixed rate per Hubble 
volume and Hubble time. This scaling property means that
the defect network looks statistically the same at any time,
once its characteristic length is normalised to $H^{-1}$.
Then, by dividing the sky cone into cells corresponding to
expansion times and Hubble volumes, one can restrict to the study of the
CMB pattern induced in Hubble--size boxes during one Hubble time.
Such studies have been performed by  Borrill et al. \cite{bo94} and
Durrer et al. \cite{du94}. In the latter paper, a
`scaling--spot--throwing' process was implemented numerically,
with CMB spots derived from a self--similar and spherically symmetric
model of texture collapse \cite{tu90,no91}.

An analytical approach to compute the multipole coefficients
${\cal C}_\ell$,  the  two--point function, and the cosmic variances 
for CMB anisotropies arising in texture models have recently been 
proposed by Magueijo \cite{ma95}.
The model allows to study texture--induced spots of arbitrary
shapes and relies on a hand--full of `free' parameters to be
specified by numerical simulations, like the number density
of spots, $\nu$, the scaling size, $d_s$, and
the brightness factor of the particular spot, $a_k$, which tells
us about its temperature relative to the mean sky temperature
(in particular, its sign indicates whether it is a hot or a cold spot).

Let us start by discussing the spots statistics. Texture 
configurations giving rise to spots in the CMB are assumed to 
arise with a constant probability per Hubble volume and Hubble time.
By dividing the whole volume in boxes of
surface $dS$ and thickness $d\ell$ we may write the
volume probability density of spot generation in the sky as follows
\begin{equation}
\label{ana1}
d{\cal P} = { \nu \over  H^{-4}} dS d\ell dt
\end{equation}

\begin{figure}[tbp]
  \begin{center}
    \leavevmode
    \epsfxsize = 9cm
    \epsfysize = 8cm
    \epsffile{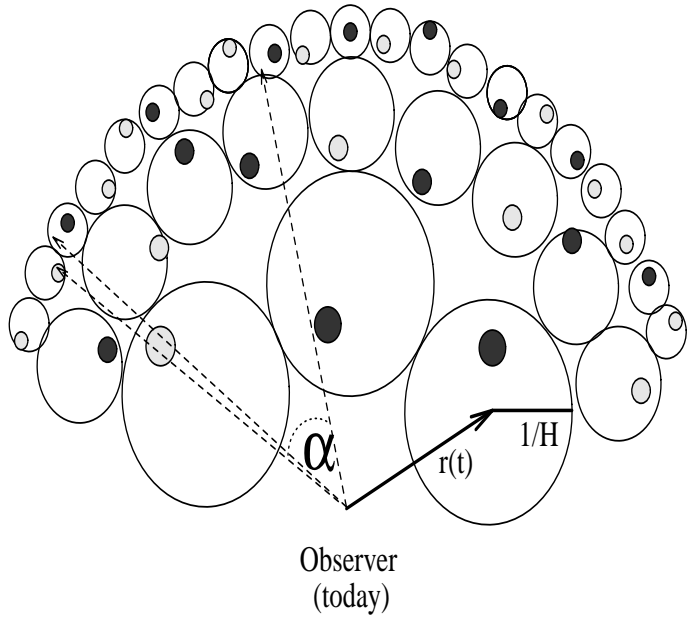}
  \end{center}
\caption{Texture--induced hot and cold spots produced on different
Hubble volumes and Hubble times, from the last scattering surface
to today (see text). Dashed arrows indicate three possible directions
on the microwave sky. The angle $\alpha$ corresponds to that in the
collapsed three--point correlation function $C_3 (\alpha )$.}
\label{figu0}
\end{figure}

where $\nu$ is the mean number of CMB spots expected to be produced 
in a Hubble volume and in a Hubble time.
Now, we are interested in a two dimensional distribution of
spots (since observations map the two dimensional sky sphere),
and thus we may integrate the thickness $d\ell$ in an interval
of order $H^{-1}$ to get the surface probability density
\begin{equation}
\label{ana1ymedio}
dP \simeq {\nu \over H^{-3} } dS dt ~.
\end{equation}

Taking into account the expansion of the universe we may express
$dS = r^2(t) d\Omega$, with
$r(t) = 3 t ((t_0 / t)^{1/3} - 1)$,
where $t$ denotes proper time and $t_0$ stands for its value today.
It is convenient to change the time variable to
$y(t) \equiv \log_{2} (t_0 / t)$, which measures how many times the
Hubble radius has doubled since time $t$ up to now
(see Fig. 1).
For a redshift $z_{\rm ls}\sim 1400$ at last scattering we have
$y_{\rm ls} \simeq \log_{2}[(1400)^{3/2}] \simeq 16$.
In terms of $y$ the surface probability density may be cast as
\begin{equation}
\label{ana2pre}
dP = N(y) dy d\Omega,
\end{equation}
with
\begin{equation}
\label{ana2}
N(y) = - {8 \nu \ln (2) \over 3} \left(2^{y/3} - 1 \right)^2 ~.
\end{equation}

Within the model, CMB anisotropies arise from
texture--induced spots in the sky.
Thus, we may express the anisotropies as the superposition of the 
contribution coming from all the individual spots $S_k$ produced from 
$y_{\rm ls}$ up to now,
\begin{equation}
{\Delta T \over T}  = \sum_{k} a_{k} S_k(\theta_k, y).
\label{DT}
\end{equation}
In this expression, $a_{k}$ describes the brightness of the hot/cold
$k$--th spot and is interpreted as a random variable, which
characteristic values have to be extracted from numerical simulations
as those of Ref. \cite{bo94}. $S_k (\theta_k, y)$ is the
characteristic shape of the spots produced at time $y$, where
$\theta_k $ is the angle in the sky measured with respect to the
center of the spot. 

A spot appearing at time $y$ has typically a size
$\theta^s(y) \simeq d_s  ~ \theta^{\rm hor}(y)$, with
$\theta^{\rm hor}(y)$ the angular size of the horizon at
$y$, and where
\begin{equation}
\label{ana3}
\theta^s(y)
= \arcsin \left( {d_s H^{-1}\over r(t)}          \right)
= \arcsin \left( {0.5 d_s   \over 2^{y/3} - 1} \right) ~.
\end{equation}

As spot anisotropies are generated by causal seeds,
their angular size cannot exceed the size of the horizon
at that time and thus  $d_s \leq 1$.
The scaling hypothesis implies that the profiles satisfy
$S_k(\theta_k , y) = S(\theta_k / \theta^s(y))$.

Combining the expression for $\Delta T /T$ in Eq. (\ref{DT}) 
and its expansion in spherical harmonics, we easily
find the expression for the multipole coefficients
$a_{\ell}^{m}$ in terms of the brightness and the profile of
the spots
\begin{equation}
\label{ana4}
a_{\ell}^{m} =
\sum_k a_k S^\ell_k (y) {Y_{\ell}^{ m}}^* (\hat\gamma _k ) ~,
\end{equation}
where
\begin{equation}
\label{ana5}
S^\ell_k(y) = 2 \pi \int_{-1}^{1} S_k(\theta_k,y) 
P_\ell (\cos\theta_k) d\cos\theta_k ~.
\end{equation}

We can now compute the angular spectrum predicted within this analytical model
\begin{equation}
\label{ana41}
{\cal C}_\ell \equiv {T_0^2 \over (2 \ell + 1)} 
\sum_{m=-\ell}^{\ell}\langle |a_{\ell}^{m}|^2 \rangle=
{T_0^2 \over (2\ell + 1)}\sum_{m=-\ell}^{\ell} \sum_{k k'} \langle a_k a_{k'}
S_k^\ell S_{k'}^\ell \rangle Y_\ell^m (\hat\gamma_k) {Y_{\ell}^{ m}}^* 
(\hat\gamma_{k'}) ~,
\end{equation}
where $\bigl\langle\cdot\bigr\rangle$ denotes ensemble average.
Any two different texture--spots are uncorrelated, hence we have
\begin{equation}
\label{ana42}
\bigl\langle a_k a_{k'} S_k^\ell S_{k'}^\ell \bigr\rangle
= \delta_{k k'} a_k^2 \left( S_k^\ell \right)^2 ~.
\end{equation}
Thus,
\begin{equation}
\label{ana43}
{\cal C}_\ell = {T_0^2\over 4\pi} \sum_k a_k^2 \left( S_k^\ell \right)^2 ~.
\end{equation}
The sum over all the spots contributing to the $\Delta T/T$ is 
performed by integrating this expression with the probability function 
in Eq. (\ref{ana2pre}), we obtain
\begin{equation}
\label{ana6}
{\cal C}_\ell = T_0^2 \bigl\langle a^2 \bigr\rangle
  {\cal I}_2^{\ell \ell }~,
\end{equation}
where we defined
\begin{equation}
\label{ana7}
{\cal I}_D^{\ell_1 \ldots \ell_D} \equiv
\int dy N(y)  S^{\ell_1}(y) \ldots  S^{\ell_D}(y) ~.
\end{equation}
$\bigl\langle  a^2 \bigr\rangle$ in Eq. (\ref{ana6}) is the mean squared value
of the spot brightness, to be computed from the distribution
\{$a_k$\} resulting from simulations.

\section{Texture three--point function}

Let us now compute the angular bispectrum. Following the same
steps as in the previous section we may express it as
\begin{equation}
\label{ana8}
\bigl\langle a_{\ell_1 }^{m_1} a_{\ell_2}^{ m_2} a_{\ell_3}^{ m_3} 
\bigr\rangle =
\bigl\langle a^3 \bigr\rangle
{\cal I}_3^{\ell_1  \ell_2  \ell_3}
\bar {\cal H}_{\ell_1 \ell_2 \ell_3}^{m_1 m_2 m_3} ~.
\end{equation}
This time the angular integral in the probability (Eq. 
(\ref{ana2pre}))
is the responsible for the appearance of the factor
$\bar {\cal H}_{\ell_1 \ell_2 \ell_3}^{m_1 m_2 m_3} $.
Replacing this expression in Eq. (\ref{coll}) we obtain the 
mean collapsed three--point function
predicted by this model
\begin{equation}
\label{collxx}
\bigl\langle C_3(\alpha) \bigr\rangle =
{\bigl\langle a^3 \bigr\rangle T_0^3\over (4\pi)^2} 
\sum_{\ell_1,\ell_2,\ell_3}
(2\ell_1 + 1) (2\ell_2 + 1) (2\ell_3 + 1) 
P_{\ell_1}(\cos\alpha)
{\cal W}_{\ell_1} {\cal W}_{\ell_2} {\cal W}_{\ell_3}
{\cal I}_3^{\ell_1 \ell_2 \ell_3}
{\cal F}_{\ell_1 \ell_2 \ell_3} ~,
\end{equation}
where ${\cal F}_{\ell_1 \ell_2 \ell_3} \equiv
\left(^{\ell_1~ \ell_2~ \ell_3}_{0~ ~0~ ~0}\right)^2$
are just the squares of the the $3j$--symbols \cite{me76}.

The other quantity of interest is the expected {\it rms} dispersion
about the three--point
function due to the cosmic variance, given by Eq. (\ref{cv}).
Computing the ensemble average
of the combination of six $a_\ell^m$'s, 
$\bigl\langle a_{\ell_1 }^{m_1} \ldots  a_{\ell_6}^{m_6}\bigr\rangle$, 
and replacing it into the expression of $\sigma_{CV}^2(\alpha)$, we 
obtain after a lengthy calculation
\begin{eqnarray}
\label{colltexturms}
\lefteqn{
\sigma_{CV}^2(\alpha )=}
\nonumber\\&&
2 ~   {\bigl\langle   a^2 \bigr\rangle^3 T_0^6\over (4\pi)^3}
\!\!\sum_{\ell_1,\ell_2,\ell_3}
\!\!P_{\ell_1}(\alpha)
( P_{\ell_1}(\alpha)\! +\! P_{\ell_2}(\alpha)\! +\! P_{\ell_3}(\alpha) )
~{\cal F}_{\ell_1 \ell_2 \ell_3}
\prod_{\ell = \ell_1}^{\ell_3}
{\cal I}_2^{\ell \ell}
{\cal W}_{\ell}^2 (2\ell + 1) 
\nonumber\\&&
+
{\bigl\langle  a^2 \bigr\rangle \bigl\langle   a^4 \bigr\rangle 
T_0^6\over (4\pi)^4}
\sum_{\ell_1,\ell_2,\ell_3,\ell_5,\ell_6}
\left[
  P_{\ell_1}^2(\alpha) +
4 P_{\ell_1}  (\alpha) P_{\ell_2}(\alpha)  +
4 P_{\ell_2}  (\alpha) P_{\ell_5}(\alpha)
\right]
\nonumber\\ &&
\times {\cal I}_2^{\ell_1 \ell_1}
{\cal I}_4^{\ell_2\ell_3\ell_5\ell_6} ~
{\cal W}^2_{\ell_1}
{\cal W}_{\ell_2}{\cal W}_{\ell_3}{\cal W}_{\ell_5}{\cal W}_{\ell_6} ~
{\cal F}_{\ell_1 \ell_2 \ell_3}{\cal F}_{\ell_1 \ell_5 \ell_6}
(2\ell_1 + 1) (2\ell_2 + 1) (2\ell_3 + 1) 
(2\ell_5+ 1) (2\ell_6 + 1) 
\nonumber\\ &&
+
{\bigl\langle  a^3 \bigr\rangle^2 T_0^6\over (4\pi)^4}
\sum_{\ell_1,\ell_2,\ell_3,\ell_5,\ell_6}
\left[
  P_{\ell_1}^2(\alpha) +
4 P_{\ell_1}  (\alpha) P_{\ell_2}(\alpha)  +
4 P_{\ell_2}  (\alpha) P_{\ell_5}(\alpha)
\right]
\nonumber\\ &&
\times {\cal W}_{\ell_1}^2
{\cal W}_{\ell_2}{\cal W}_{\ell_3}
{\cal W}_{\ell_5}{\cal W}_{\ell_6}
{\cal I}_3^{\ell_1\ell_2\ell_3}
{\cal I}_3^{\ell_1\ell_5\ell_6}
{\cal F}_{\ell_1 \ell_2 \ell_3}
{\cal F}_{\ell_1 \ell_5 \ell_6}
(2\ell_1 + 1) (2\ell_2 + 1) (2\ell_3 + 1) 
(2\ell_5 + 1) (2\ell_6 + 1) 
\nonumber\\ &&
+
{\bigl\langle  a^6 \bigr\rangle T_0^6\over(4\pi)^5}
\sum_{\ell_1,\ell_2,\ell_3,\ell_4,\ell_5,\ell_6}
P_{\ell_1}(\alpha) P_{\ell_4}(\alpha) ~
{\cal W}_{\ell_1}{\cal W}_{\ell_2}{\cal W}_{\ell_3}
{\cal W}_{\ell_4}{\cal W}_{\ell_5}{\cal W}_{\ell_6}
{\cal I}_6^{\ell_1\ell_2\ell_3\ell_4\ell_5\ell_6}
\nonumber\\ &&
\times
{\cal F}_{\ell_1 \ell_2 \ell_3}{\cal F}_{\ell_4 \ell_5 \ell_6}
(2\ell_1 + 1) (2\ell_2 + 1) (2\ell_3 + 1) 
(2\ell_4 + 1) (2\ell_5 + 1) (2\ell_6 + 1) 
~.
\end{eqnarray}

We can estimate the range for the amplitude of the 
three--point correlation function predicted by the model by 
$\bigl\langle  C_3 (\alpha ) \bigr\rangle \pm \sigma_{CV}(\alpha )$.

Let us  note that a brightness distribution \{$a_k$\} 
symmetric in hot and cold spots implies  $\bigl\langle 
a^3 \bigr\rangle = 0$ and therefore yields a vanishing
three--point function.
This is precisely what happens for spots
generated by spherically symmetric self--similar (SSSS) texture
unwinding events \cite{tu90}.
However, the properties of the
SSSS solution are not characteristic of spot arising from more realistic
random configurations.
Borrill et al. \cite{bo94} have shown that randomly generated texture
field configurations produce anisotropy patterns with
very different properties to the exact SSSS solution.
Furthermore, spots generated from concentrations of energy gradients
which do not lead to unwinding events can still produce
anisotropies very similar to those generated by unwindings. 
The peak anisotropy of the random configurations turns
out to be 20 to 40 \% smaller than the SSSS
solution. Moreover, their results also suggest an asymmetry
between maxima $\langle a_{\rm max} \rangle$ and minima 
$\langle a_{\rm min} \rangle$ of the peaks for all studied
configurations other than the SSSS, although with large error bars.
They justify this result as due to the fact that,
for unwinding events, the minima are
generated earlier in the evolution (photons climbing out of the
collapsing texture) than the maxima (photons falling in the collapsing
texture), and thus the field correlations are stronger for the maxima,
which enhance the anisotropies.

\section{Numerical results}

Up to this point we have left unspecified the profile of the spots, the
distribution of the brightness factor $\{a_k\}$, the typical size of
the spots $d_s$ and the number density of spots $\nu$. We have to fix 
them in order to obtain a quantitative estimate of the three--point
function and the cosmic variance. We will essentially rely on the
results of the simulations of Ref. \cite{bo94}, where all these issues
are addressed to some extent.
These simulations indicate that the size of the spots are
approximately 10\% of the Hubble radius when generated, hence we take
the scaling size $d_s \simeq 0.1$.
There is no particular form of the spot profiles
obtained in the simulations. However,
a Gaussian function, as given by
the profile  $S(\theta / \theta_s(y)) \propto
\exp (- \theta^2 / \theta^2_s(y))$ 
looks as a good approximation to the bell--shaped spots obtained
in Ref. \cite{bo94} and we will adopt it here for the numerical 
computations. Another interesting result obtained in Ref. \cite{bo94} 
regards the number of spots produced: whereas 
only 4 out of 100 simulations showed unwinding events,
the number density of nonunwinding events (which as we mentioned
above also lead to spots) is much greater, typically of the
order of one such event per simulation.
This implies a mean number of CMB spots per Hubble volume
and Hubble time $\nu \simeq 1$.

We need also the distribution of the spot brightness $\{a_k\}$ in order to
compute the mean values of the powers of the random variable $a$, 
$\langle a^n \rangle$,  appearing 
in Eqs. (\ref{collxx}) and (\ref{colltexturms}). As we do not have 
at our disposal the full distribution of values obtained in the
simulations and we just know the mean value of the maxima and the
minima, we will make a crude estimate by considering that 
all the hot spots appear with the same $a_h > 0$ and all the cold 
spots with the same $a_c < 0$. 
Then, all the $\langle a^n \rangle$ 
needed can readily be obtained in terms of $\langle a^2\rangle$ and 
$x\equiv \langle a \rangle /\langle |a| \rangle$. From the two--point 
correlation function of the {\sl COBE}--DMR data, we can normalize the 
amplitude of the anisotropies and fix $\langle a^2 \rangle$. We do 
this by fixing ${\cal C}_9=(8 \mu {\rm K})^2$ \cite{go94} which, with the
help of Eq. (\ref{ana6}), leads to 
$\langle a^2 \rangle (T_0)^2 = 1.4 \times 10^5 \mu {\rm K}^2$. 
The other
parameter, $x$, measures the possible {\sl asymmetry} between hot and cold 
spots and we leave it as a free parameter. Let us note that the 
amplitude of the spots is directly related to the scale of symmetry 
breaking. From Ref. \cite{bo94}, $\langle a^2\rangle^{1/2} \simeq 0.7 
\epsilon$, with $\epsilon= 8\pi^2 G \Phi_0^2$. Hence, the symmetry 
breaking scale results $\Phi_0 \simeq 1.6 \times 10^{-3} m_P$, typical of GUT 
theories.

We may now apply our formalism to the {\sl COBE}--DMR measurements. We 
consider a Gaussian window function, ${\cal W}_\ell \simeq 
\exp(-\ell(\ell+1)\sigma^2/2)$, with dispersion of the antenna beam 
profile $\sigma=3.2^\circ$ \cite{wr92}.  
The analytical formulae obtained in the previous section assumed 
full--sky coverage.
The effect of partial coverage, due to the cut in the maps at Galactic
latitudes $\vert b \vert < 20^\circ$, increases the cosmic variance
by a factor $\sim 1.56$ \cite{hi94}, and we take it into account by 
multiplying $\sigma_{CV}$ by this factor in the numerical results.
\begin{figure}[tbp]
  \begin{center}
    \leavevmode
    \epsfxsize = 9cm
    \epsfysize = 8cm
    \epsffile{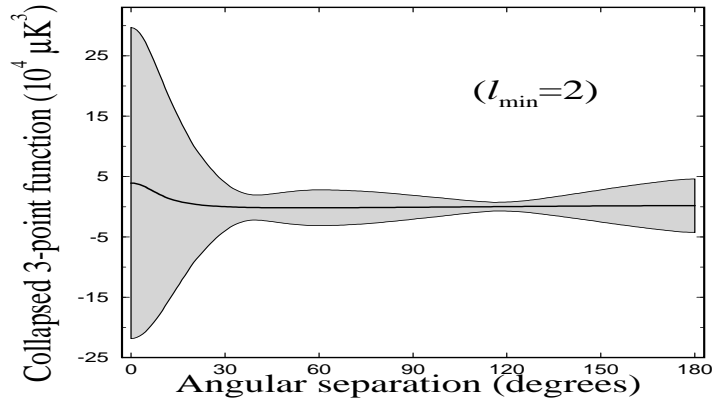}
  \end{center}
\vspace{-2.0cm}
\caption{Collapsed three--point function
$\bigl\langle C_3 (\alpha ) \bigr\rangle$ (solid line) and 
the {\it rms} cosmic variance (grey band), 
in $10^4~\mu {\rm K}^3$ units, 
arising from the 
texture--induced temperature fluctuations.
The band includes the $\sim 50\%$ increment in $\sigma_{CV}$ due to
the sample variance.
The asymmetry parameter is $x=0.07$ and 
the dipole contribution is subtracted.}
\label{figu1}
\end{figure}
Figure 2 shows the collapsed three--point function
$\bigl\langle C_3 (\alpha ) \bigr\rangle$  (solid line) and the grey 
band indicates the {\it rms} range of fluctuations expected from the 
cosmic variance. We have plotted the results for $x=0.07$, 
which looks as a reasonable value according to Ref. \cite{bo94}. It is 
easy to find how these results change for other values of the parameter 
$x$, as the central curve $\langle C_3(\alpha)\rangle \propto x$, 
while the variance $\sigma_{CV}$ is approximately constant for small 
values of the $x$ parameter. 
The results do not include the dipole $(\ell_{min} =2)$, 
to match the {\sl COBE} analysis. These results 
are to be compared with the data in Ref. \cite{hi95}. For all 
reasonable values of the parameter $x$, the data falls well within the 
grey band, and thus there is good agreement with the observations. The 
cosmic variance band coming from the texture model is significantly 
larger than that coming from Gaussian distributed anisotropies 
\cite{hi95,mo95}. It is more than a factor 10 larger for angles smaller 
than $20^\circ$ and a factor 5 larger for larger angles. 
An increment in the cosmic variance has also been pointed out in 
Ref. \cite{ma95} for the two--point correlation function, and has more
extensively been studied in Ref. \cite{ma95b}. 
The fact that the range of 
expected values for the three--point correlation function predicted by 
inflation is included into that predicted by textures for all the 
angles and that the data points fall within them makes it impossible to 
draw conclusions favouring one of the models, although one may think that 
the fact that the data covers the full range of the band predicted by 
inflation means that the observed sky is a typical realization of the 
ensemble, while it is a less probable realization in the texture model 
as all the data fall close to the origin and the wide range of 
values allowed by the cosmic variance is nearly void.

The largest contribution to the cosmic variance comes from the small 
values of $\ell$. Thus, the situation may improve if one subtracts
the lower order multipoles contribution, as in the 
$\ell_{min}=10$ analysis of Ref. \cite{hi95}. 
\begin{figure}[tbp]
\begin{center}
\leavevmode
{\hbox %
{\epsfxsize = 7cm \epsffile{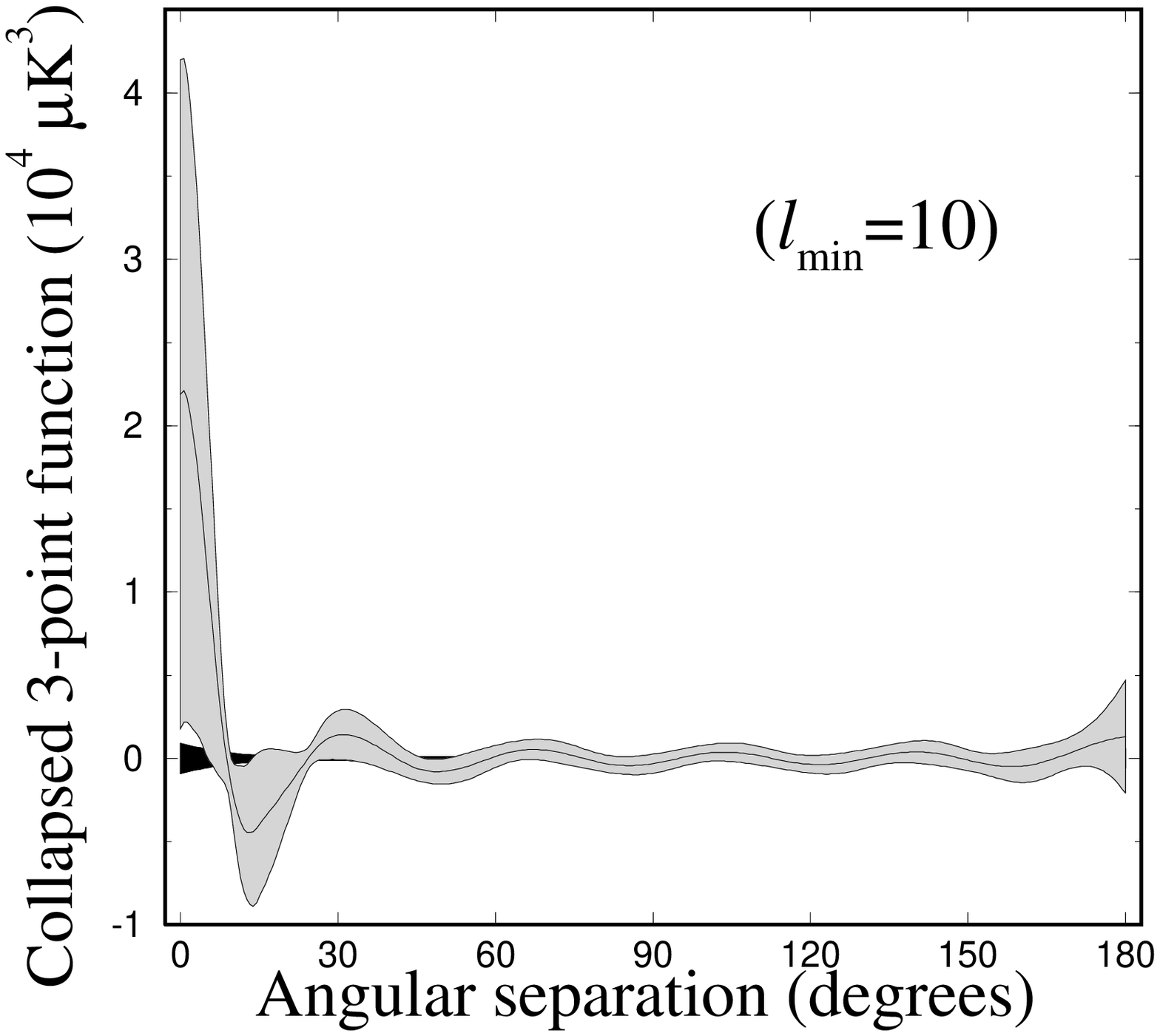} }
{$~~$}
{\epsfxsize = 7cm \epsffile{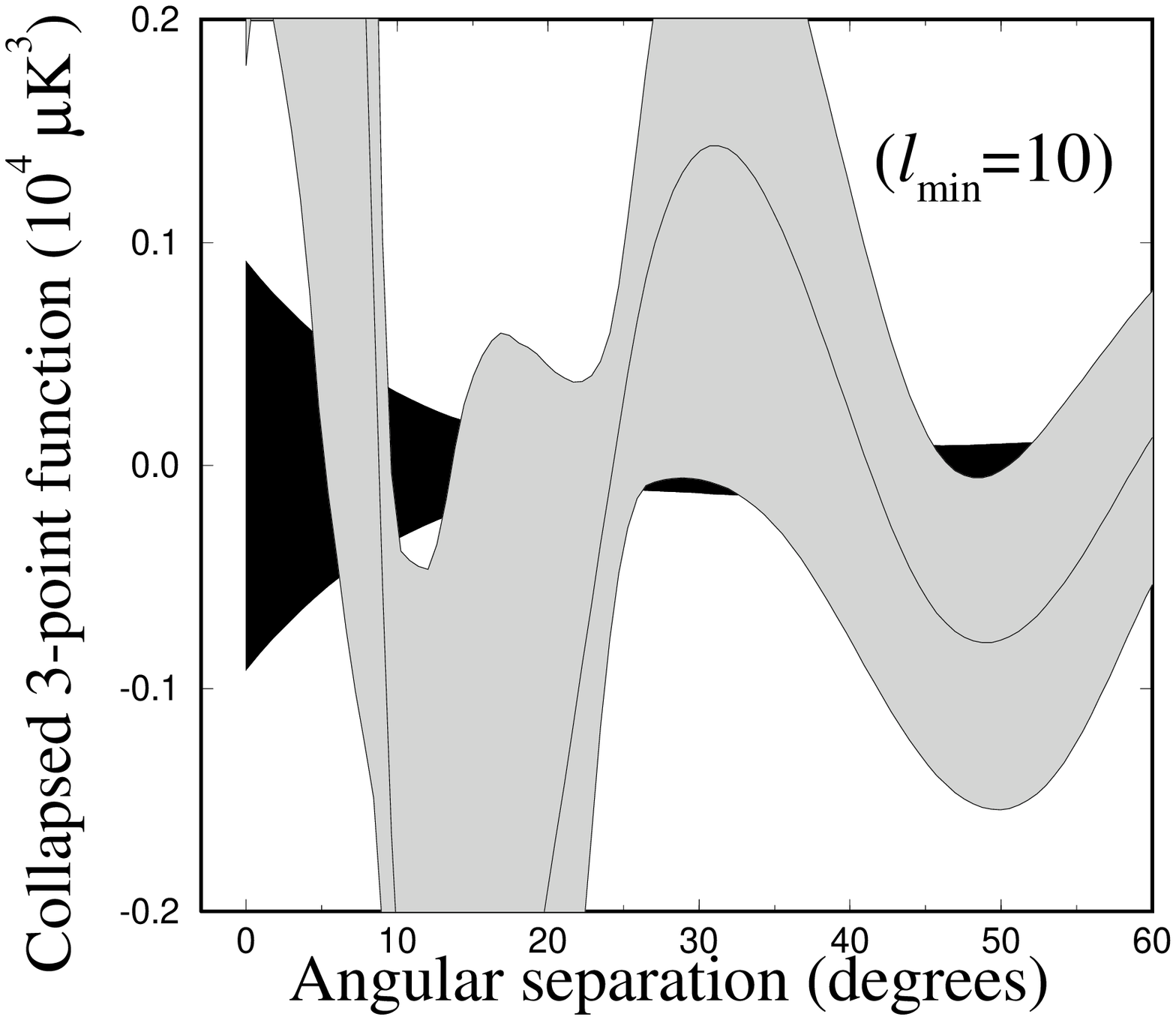} } }
\end{center}
\vspace{-2.0cm}
\caption{Expected range of values for the three--point correlation
function for the texture model 
with asymmetry parameter $x=0.4$
(grey band, as in Fig. 2).
Also shown is the expected range for inflationary models (black band).
Both bands include the $\sim 50\%$ increase in $\sigma_{CV}$ due to
the sample variance.
All multipoles up to $\ell = 9$ have been subtracted.
The right panel 
shows a zoomed fraction of the same plot.}
\label{Figu2a-b}
\end{figure}
In Figures 3 a--b, we show by a grey 
band the expected range of values for the three--point correlation 
function for the texture model (as in Fig. 2) and by a black band the 
expected range for inflationary models. The bands do not superpose each 
other for some ranges of values of the separation angle for the value 
of $x=0.4$ considered, what means that measurements in that range can 
distinguish among the models. The value of the parameter $x$ 
considered is quite large, much in excess of that suggested by Ref. 
\cite{bo94}, but we have chosen it to show an example with a 
noticeable effect. We have checked that the non--superposition of the 
bands appears for $|x|>0.33$. The comparison with the data is less 
clear for this case, as the two--year {\sl COBE} results have large 
instrumental noise (in excess of the Gaussian cosmic variance) for 
this subtraction scheme \cite{hi95}. However, with the four--year 
data, the instrumental noise is expected to diminish under the cosmic 
variance, and it would be possible to distinguish among the models, at 
least for textures with $|x|>0.33$. On the other hand, already from 
the present data of Ref. \cite{hi95} for small and very large angles, 
we can say that models with $x< -0.4$ are quite disfavoured.

An experiment probing smaller angular scales than {\sl COBE} should thus 
be more appropriate to test non--Gaussian features in texture models. 
As an example we compute the predictions for a three--beam subtraction 
scheme experiment with window function at zero--lag
${\cal W}_\ell =
(1.5 - 2 P_\ell (\cos\theta) + 0.5 P_\ell (\cos 2\theta))^{1/2}
\exp(-\ell(\ell+1)\sigma^2/2)$,
where $\sigma = 0.64^\circ$ is the beam width and $\theta = 2.57^\circ$
is the chopping angle. 
This window function is peaked at $\ell \sim 70$ and the range 
of multipoles that significantly contribute to the three--point 
function is from $\ell=10$ to $\ell=100$. 
Hence we are still probing large enough scales and our results
are not strongly affected by the microphysics of the last scattering surface, 
as it has recently been shown that the scale for the appearance 
of the so--called Doppler peaks within texture models is shifted  
to larger $\ell$'s than in the standard adiabatic case and their
contribution turns out not to be dominant for this range of $\ell$'s
\cite{ct95,dgs96}.

We obtain for the skewness 
$S\equiv \langle C_3(0)\rangle=(4.75 \pm 1.94) \times 10^4 \mu{\rm K}^3$,
where the error band stands for the associated cosmic variance 
$\sigma_{CV}^2(0)$, 
for a value of the asymmetry parameter $x=0.07$.
For comparison, the Gaussian adiabatic prediction is 
$S_{\rm Gauss}= 0.06 \times 10^4 \mu{\rm K}^3$.
Thus, in this case even for reasonably small values of the asymmetry
parameter $x$ one such experiment can in principle distinguish between 
inflation and texture predictions, 
and thus put stronger constraints on the model parameters.

\section{Discussion}

We used an analytical model of the CMB anisotropies produced by 
textures to estimate the three--point correlation function and
its cosmic variance. In spite of its simplicity, we expect that it can
give a reasonable estimation of the results that would be obtained in
numerical simulations. The idea of this analytical model \cite{ma95}
is that the texture evolution generates hot and cold spots in the sky
at a constant rate per Hubble volume and Hubble time. Due to the
scaling property of textures evolution the typical size of the spots 
is a constant fraction of the angle subtended by the Hubble radius at 
the time they are produced. The spot shapes are
given by a profile function times a brightness factor.
These free parameters have to be determined from the results of
simulations. Here we have fixed the number density of spots produced per
Hubble volume and Hubble time, the angular scale of the spots and the
spots profile so that they are consistent with the results of
Ref. \cite{bo94}. The computations of the three--point function and the
cosmic variance are fundamentally sensitive to the distribution of the
spots brightness. A distribution symmetric in hot and cold spots
gives rise to a vanishing three--point function, while a
non--symmetric distribution yields a non--vanishing one. 
For definiteness, we parameterized the distribution with a very simple
ansatz with just two free parameters, namely 
its dispersion, $\langle a^2 \rangle$, 
and a parameter measuring the asymmetry of the distribution, $x$.
$\langle a^2 \rangle$ was fixed using the measured amplitude of the 
two--point correlation function. $x$ instead was left free and encodes the 
strongest dependence of our results on the model parameters.
As discussed above, numerical simulations for
random initial conditions point out to a non--vanishing asymmetry,
although with a small value for $x$ (and large error bars). 
It is straightforward to extend the calculations to another brightness 
distribution. 

The application of this analysis to the {\sl COBE}--DMR experimental
setting shows a very large cosmic variance band, much larger than that 
associated to a Gaussian process. For all reasonable values of
$x$, this band encloses both the Gaussian one and the COBE data, 
and thus no conclusions favouring inflationary or texture theories can be
obtained. 
The predictions of both models are consistent with the three--point
function of the {\sl COBE}--DMR maps.
On the other hand when all the multipoles up to $\ell=9$ are 
subtracted, the situation looks a bit more promising and we saw that for 
large values of $|x|$, with the four years {\sl COBE} data we could 
distinguish among texture and inflationary models. 
However, these large values of $x$ seem to be larger than those expected 
from simulations. 
Finally, we showed that a experiment probing smaller angular scales 
should be able to distinguish between both classes of models, even for 
small values of the modulus of the parameter $x$.
This looks as an encouraging result in the task of testing 
theories of the primordial fluctuations. 

\acknowledgements

The Italian MURST is acknowledged for financial support. 
A.G. acknowledges partial funding from The British Council/Fundaci\'on 
Antorchas, 
and thanks John Barrow, Andrew Liddle and the cosmology group at the 
University of Sussex for hospitality while part of this work was carried out.

\end{document}